\documentclass{article}
\usepackage{graphicx}
\usepackage{amsmath}

\input{tcilatex}

\begin{document}

\title{Note on Invariants of the Weyl Tensor}
\author{Bogdan Ni\c{t}\u{a}\thanks{%
E-mail: bnita@utdallas.edu} and Ivor Robinson\thanks{%
E-mail: robinson@utdallas.edu} \\
Department of Mathematical Sciences, EC 35,\\
University of Texas at Dallas, PO BOX 830688,\\
Richardson, TX 75083-0688 USA}
\maketitle

\begin{abstract}
Algebraically special gravitational fields are described using algebraic and
differential invariants of the Weyl tensor. A type III invariant is also
given and calculated for Robinson-Trautman spaces.

\begin{description}
\item[Key words]  : invariants, algebraic classification of the Weyl tensor.
\end{description}
\end{abstract}

\section{Introduction}

It is well known (see [1] and [2]) that the two algebraic invariants 
\begin{eqnarray}
I &=&\frac{1}{4}{}^{+}C_{abcd}{}^{+}C^{abcd}, \\
J &=&{}\frac{1}{8}{}^{+}C_{abcd}{}^{+}C^{cd}{}_{ef}{}^{+}C^{efab},
\end{eqnarray}
where $^{+}C$ is the self-dual part of the Weyl tensor, provide a partial
classification of the Weyl tensor. Moreover, if $\chi $ is the cross ratio
of any four null directions then 
\begin{equation}
I^{3}\left[ \left( \chi +1\right) \left( \chi -2\right) \left( \chi -\frac{1%
}{2}\right) \right] ^{2}=6J^{2}\left[ \left( \chi +\omega \right) \left(
\chi +\omega ^{2}\right) \right] ^{3}
\end{equation}
where $\omega =e^{\frac{2\pi i}{3}}.$ In particular $I^{3}=6J^{2}\neq 0$ $%
\Leftrightarrow $ $\chi \in \left\{ 0,1,\infty \right\} $ $%
\Longleftrightarrow $ (2,1,1) or (2,2).

\section{Classification}

For any $F_{abcd}$ with symmetries similar to the ones of $^{+}C$ we define 
\begin{equation}
J_{F}=F_{abcd;rs}F^{abcd}{}_{;tu}\overline{F}_{efgh;}{}^{rs}\overline{F}%
^{efgh;tu};
\end{equation}
remark that for any null field $F,$ $J_{F}$ is the invariant $J$ in [3]. We
are particularly interested in $J_{A},$ $J_{B}$ and $J_{^{+}C}$ where 
\begin{eqnarray}
A_{abcd} &=&IB_{abcd}-J^{+}C_{abcd} \\
B^{ab}{}_{cd} &=&\frac{1}{2}{}^{+}C^{ab}{}_{rs}{}^{+}C^{sr}{}_{cd}-\frac{1}{3%
}I{}^{+}\delta _{cd}^{ab}
\end{eqnarray}
where $\delta _{cd}^{ab}=\frac{1}{2}\left( g_{ad}g_{bc}-g_{ac}g_{bd}-i\eta
_{abcd}\right) ,$ $\eta $ being the Levi-Civita tensor. Notice that when $%
I^{3}=6J^{2}$ the tensor $A$ is null ($A_{abcd}=6\Psi _{2}^{2}(3\Psi
_{2}\Psi _{4}-\Psi _{3}^{2})N_{ab}N_{cd})$ in the case (2,1,1) and it
vanishes in the more degenerate cases (see [1]); for $I=J=0$ the tensor $B$
is null ($B_{abcd}=-4\Psi _{3}^{2}N_{ab}N_{cd})$ in the (3,1) case and zero
otherwise. Moreover 
\begin{eqnarray}
J_{A} &=&\left| 96\Psi _{2}^{2}(3\Psi _{2}\Psi _{4}-\Psi _{3}^{2})\rho
^{2}\right| ^{4}, \\
J_{B} &=&\left| 8\Psi _{3}\rho \right| ^{8}.
\end{eqnarray}

In conclusion, for space-times admitting an expanding congruence we have the
following classification:

- $I^{3}\neq 6J^{2},$ $I\neq 0,$ $J\neq 0:$ (1,1,1,1);

- $I^{3}=6J^{2}\neq 0$, $J_{A}\neq 0:$ (2,1,1);

- $I^{3}=6J^{2}\neq 0$, $J_{A}=0:$ (2,2);

- $I=J=0,$ $J_{B}\neq 0:$ (3,1);

- $I=J=0,$ $J_{B}=0,$ $J_{^{+}C}\neq 0:$ (4);

- $I=J=0,$ $J_{B}=0,$ $J_{^{+}C}=0:$ (-).

\section{Further remarks on the (3,1) case}

For $I=J=0$ case we can alternatively use the first order invariant obtained
in [4] 
\begin{equation}
J_{P}=C^{abcd;e}C_{amcn;e}C^{lmrn;s}C_{lbrd;s}
\end{equation}
to distinct (3,1) case from more degenerate ones.

We did not investigate systematically invariants of second order but we
mention that if 
\begin{equation}
D_{rst}={}^{+}C_{abcd;r}{}^{+}C^{abcd}{}_{;st}
\end{equation}
then 
\begin{equation}
D=D_{\left[ rs\right] t}\overline{D}^{\left[ rs\right] t}
\end{equation}
has the following expression for a (3,1)\ Robinson-Trautman solution with $%
P=P(\sigma ,\xi ,\eta ):$%
\begin{eqnarray}
D &=&\frac{36p^{4}}{r^{14}}\left( K_{\xi }^{2}+K_{\eta }^{2}\right) \left[ 
\frac{1}{8}\left( K_{\xi }^{2}+K_{\eta }^{2}\right) K+p\left( K_{\xi \eta
}^{2}-K_{\xi \xi }K_{\eta \eta }\right) \right] \\
&&+9\frac{p^{4}}{r^{13}}\left[ \left( K_{\xi }^{2}+K_{\eta }^{2}\right) ^{2}%
\right] ,_{\sigma }.  \notag
\end{eqnarray}

Remark that for (3,1) and (4) cases, the geometry of each light cone is
independent of the one of its neighbors; and both $J_{P}$ and $J_{B}$ depend
only on the geometry of each individual light cone. However, the invariant $%
D $ also depends on the rate of change of the geometry from one light cone
to another.

\bigskip

\noindent \textbf{References}

\bigskip

\noindent \lbrack 1] Peres A ''Invariants of General Relativity II -
Classification of spaces'' \emph{Il Nuovo Cimento} \textbf{18} (1960) 36.

\noindent \lbrack 2] Penrose R and Rindler W ''Spinors and Space-time Vol.
2'' (Cambridge: Cambridge University Press).

\noindent \lbrack 3] Ni\c{t}\u{a} B and Robinson I ''An Invariant of Null
Spinor Fields'' \emph{Class. Quantum Grav.} \textbf{17} (2000) 2153.

\noindent \lbrack 4] Pravda V ''Curvature invariants in type III
space-times'' \emph{Class. Quantum Grav.} \textbf{16} (1999) 3321.

\end{document}